\documentclass[10pt,conference,letterpaper,nofonttune]{IEEEtran}
\IEEEoverridecommandlockouts

\usepackage{cite}
\usepackage{amsmath,amssymb,amsfonts}
\usepackage{algorithmic}
\usepackage{graphicx}
\usepackage{textcomp}
\usepackage{xcolor}
\usepackage{fancyhdr}
\fancyheadoffset[L]{1cm}

\usepackage{acronym}
\usepackage{bm}
\usepackage{subcaption}
\usepackage{adjustbox}
\usepackage{multirow}
\usepackage{multicol}
\usepackage{balance}
\usepackage{soul}

\usepackage{booktabs}
\usepackage{array}
\usepackage[font=footnotesize]{subcaption}
\usepackage[font=footnotesize]{caption}

\acrodef{5G NR}[5G NR]{5G New Radio}
\acrodef{AMR}[AMR]{Autonomous Mobile Robot}
\acrodef{AoI}[AoI]{Age of Information}
\acrodef{CCDF}[CCDF]{Complementary Cumulative Distribution Function}
\acrodef{CU}[CU]{Central Unit}
\acrodef{CN}[CN]{Core Network}
\acrodef{DU}[DU]{Distributed Unit}
\acrodef{FOV}[FOV]{Field Of View}
\acrodef{IIoT}[IIoT]{Industrial Internet of Things}
\acrodef{KPI}[KPI]{Key Performance Indicator}
\acrodef{near-RT}[near-RT]{near-real-time}
\acrodef{non-RT}[non-RT]{non-real-time}
\acrodef{MAC}[MAC]{Medium Access Control}
\acrodef{MCHEM}[MCHEM]{Massive Channel Emulator}
\acrodef{OAI}[OAI]{OpenAirInterface}
\acrodef{OSC}[OSC]{O-RAN Software Community}
\acrodef{PDCP}[PDCP]{Packet Data Convergence Protocol}
\acrodef{PDU}[PDU]{Protocol Data Unit}
\acrodef{PF}[PF]{Proportional Fair}
\acrodef{QoS}[QoS]{Quality of Service}
\acrodef{RAN}[RAN]{Radio Access Network}
\acrodef{RB}[RB]{Resource Block}
\acrodef{RIC}[RIC]{RAN Intelligent Controller}
\acrodef{RU}[RU]{Radio Unit}
\acrodef{SDR}[SDR]{Software Defined Radio}
\acrodef{SM}[SM]{Service Model}
\acrodef{SMO}[SMO]{Service Management and Orchestration}
\acrodef{SRN}[SRN]{Standard Radio Node}
\acrodef{TB}[TB]{Transport Block}
\acrodef{TBS}[TBS]{Transport Block Size}
\acrodef{UE}[UE]{User Equipment}
\acrodef{UPF}[UPF]{User Plane Function}
\acrodef{URLLC}[URLLC]{Ultra Reliable Low Latency Communication}

\definecolor{mygreen}{RGB}{34, 194, 2} 

\title{TailO-RAN: O-RAN Control on Scheduler Parameters to Tailor RAN Performance}

\author{
    \IEEEauthorblockN{ 
        Nicolò~Longhi\IEEEauthorrefmark{1}\IEEEauthorrefmark{2},
        Salvatore~D'Oro\IEEEauthorrefmark{3},
        Leonardo~Bonati\IEEEauthorrefmark{3}, 
        Michele~Polese\IEEEauthorrefmark{3},\\
        Roberto~Verdone\IEEEauthorrefmark{1}\IEEEauthorrefmark{2},
        and Tommaso~Melodia\IEEEauthorrefmark{3}
    }
    \IEEEauthorblockA{\IEEEauthorrefmark{1} Department of Electrical, Electronics, and Information Engineering, University of Bologna, Italy}
    \IEEEauthorblockA{\IEEEauthorrefmark{2} WiLab - National Wireless Communication Laboratory (CNIT), Bologna, Italy}
     \IEEEauthorblockA{\IEEEauthorrefmark{3} Institute for the Wireless Internet of Things, Northeastern University, Boston, MA, U.S.A.}
}

\begin{document}

\bstctlcite{IEEEexample:BSTcontrol}

\maketitle

\thispagestyle{fancy}
\fancyhead[L]{This paper has been accepted for publication on IEEE Global Communications Conference (GLOBECOM) 2025.\\ \scriptsize \textcopyright 2025 IEEE.Personal use of this material is permitted.  Permission from IEEE must be obtained for all other uses, in any current or future media, including reprinting/republishing this material for advertising or promotional purposes, creating new collective works, for resale or redistribution to servers or lists, or reuse of any copyrighted component of this work in other works}

\begin{abstract}
The traditional black-box and monolithic approach to \acp{RAN} has heavily limited flexibility and innovation. The Open RAN paradigm, and the architecture proposed by the O-RAN ALLIANCE, aim to address these limitations via openness, virtualization and network intelligence. 
%
In this work, first we propose a novel, programmable scheduler design for Open RAN \acp{DU} that can guarantee minimum throughput levels to \acp{UE} via configurable weights. Then, we propose an O-RAN xApp that reconfigures the scheduler's weights dynamically based on the joint \ac{CCDF} of reported throughput values. We demonstrate the effectiveness of our approach by considering the problem of asset tracking in 5G-powered \ac{IIoT} where uplink video transmissions from a set of cameras are used to detect and track assets via computer vision algorithms. 
%
%
We implement our programmable scheduler on the \ac{OAI} 5G protocol stack, and test the effectiveness of our xApp control by deploying it on the \ac{OSC} near-RT \ac{RIC} and controlling a 5G \ac{RAN} instantiated on the Colosseum Open \ac{RAN} digital twin.
%
Our experimental results demonstrate that our approach enhances the success percentage of meeting throughput requirements by 33\% compared to a reference scheduler. Moreover, in the asset tracking use case, we show that the xApp improves the detection accuracy, i.e., the F1 score,  by up to 37.04\%.

    
    \acresetall
\end{abstract}


\acresetall

\section{Introduction}

In recent years, 5G has experienced the emergence of increasingly demanding applications, a trend that will intensify with the advent of 6G. One of the most demanding verticals in terms of latency and reliability is \ac{IIoT}~\cite{3gpp.TS.22.104}. This is a new 5G use case that involves the use of 5G-connected devices, sensors, and robots in industrial environments related to manufacturing, logistics, and energy. \ac{IIoT} enables real-time monitoring, automation, and optimization of industrial processes, which requires \ac{URLLC} to ensure smooth operations and safety of industrial plants and systems. This demand, however, is not easy to meet due to the current state of cellular networks, which still heavily rely on closed and inflexible architectures. Indeed, these closed architectures can hardly dynamically and algorithmically adapt to changing network demands, or support the integration of new technologies. This limits innovation and complicates guaranteeing the minimum performance requirements needed by demanding applications. 

The Open RAN paradigm, and specifically O-RAN, addresses these limitations by promoting openness, disaggregation, virtualization, and introducing intelligence into the network \cite{10024837}. 
O-RAN implements the 7.2x split architecture, partitioning the \ac{RAN} into three nodes: \ac{CU}, \ac{DU}, and \ac{RU}. Additionally, O-RAN enables network operators to enjoy greater flexibility and gain full control of the network via the so-called \acp{RIC}. These allow operators to tailor the \ac{RAN} to specific application needs and requirements using algorithms that execute in real-time.
The two \acp{RIC} introduced by the O-RAN architecture are the \ac{near-RT} \ac{RIC} and the \ac{non-RT} \ac{RIC}. The \ac{near-RT} \ac{RIC} supports control loops with a response time between $10$\:ms and $1$\:s and is connected to the \ac{CU} and \ac{DU} through the E2 interface, while the \ac{non-RT} \ac{RIC} operates with control loops greater than $1$\:s and functions as part of the \ac{SMO} framework. The intelligence in the \ac{near-RT} \ac{RIC} is executed by applications known as xApps, while rApps perform this function in the \ac{non-RT} \ac{RIC}. 
To optimize the \ac{RAN} through such applications, the gNB nodes must provide the data required by the xApp/rApp and expose control ``knobs'' that can be tuned.

Despite its strict \ac{QoS} requirements, \ac{IIoT} traffic often exhibits spatio-temporal correlations~\cite{3gpp.TS.22.104, 5g-acia}, 
that can be used to reconfigure the \ac{RAN}, for example optimizing \ac{RAN} parameters related to pre-scheduling or \ac{PDCP} duplication~\cite{O-RAN-WG1-UCAR-r3-v14}. Because of this, \ac{IIoT} is an ideal candidate to benefit from O-RAN data-driven intelligence, as operators can dynamically tailor \ac{RAN} performance to the specific applications being served~\cite{10143032}.

Several studies have investigated how O-RAN can be applied to \ac{IIoT} scenarios. In particular, the authors of \cite{10620233} propose an O-RAN-based \ac{URLLC} scheduler that involves both non-real-time and real-time control loops to meet \ac{IIoT} requirements. Additionally, \cite{9813589} presents an O-RAN based solution for managing slicing in an \ac{IIoT} scenario, aiming to minimize \ac{AoI} while considering constraints on slice isolation and energy consumption. Concerning end-to-end experimental research, \cite{10279730} describes a testbed with proactive resource allocation by predicting \ac{UE} traffic.

In this paper, we advance the state-of-the-art by proposing a novel scheduler whose logic can be tuned to guarantee a minimum throughput to each \ac{UE}.
We show how the proposed scheduler can be dynamically controlled and programmed in an O-RAN deployment via xApps that leverage spatiotemporal correlations in traffic patterns to adapt scheduling policies based on application layer requirements. 
Although similar topics have been explored in the literature, we are the first to jointly address these aspects and validate our proposed solution on an end-to-end experimental testbed. 

\begin{figure}[t!]
  \centering
  \vspace{-2mm}
  \begin{subfigure}{0.18\textwidth}
    \includegraphics[width=\linewidth]{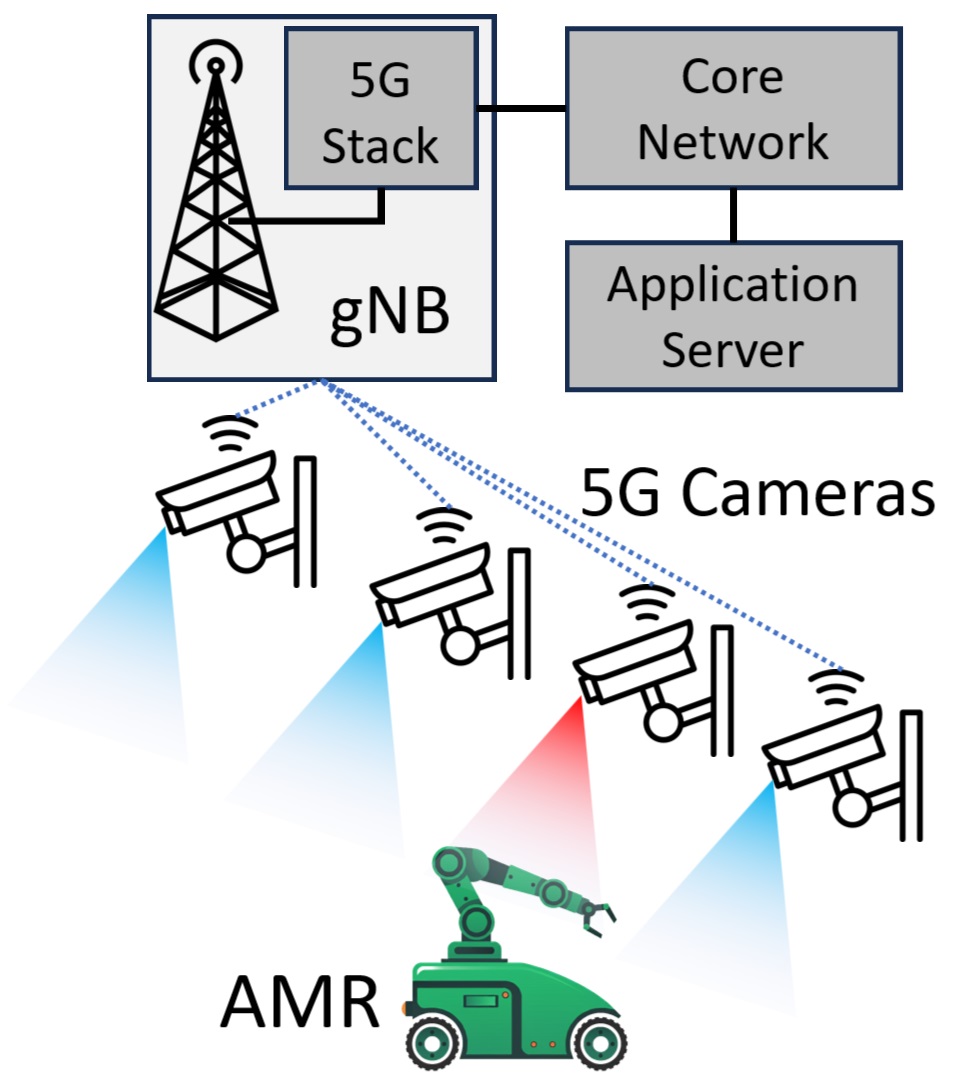}
    \caption{Use case representation}
  \end{subfigure}%
  \hspace{0.05\textwidth}
  \begin{subfigure}{0.17\textwidth}
    \includegraphics[width=0.87\linewidth]{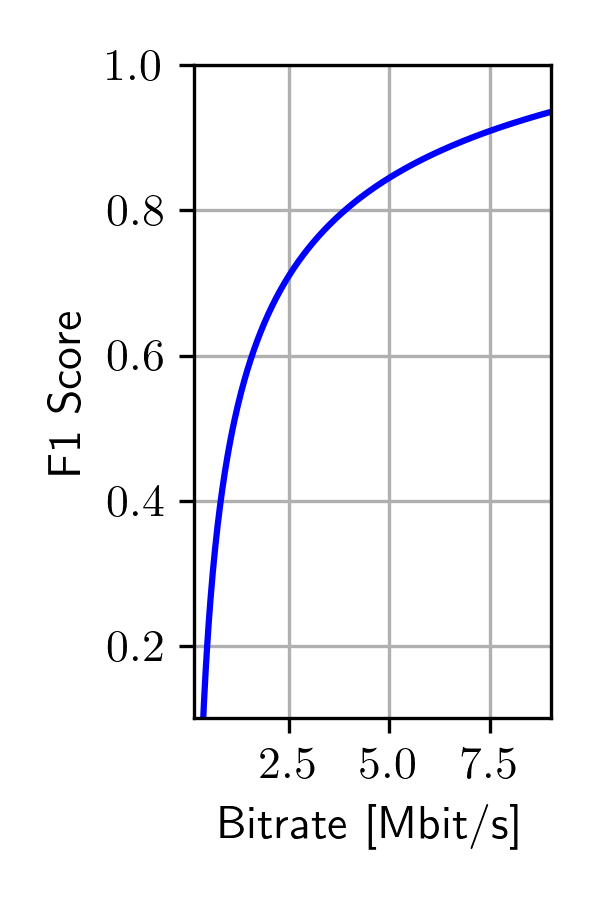}
    \caption{F1 score vs video bitrate}
  \end{subfigure}%
  \caption{Representation of the use case. The first figure shows four cameras and their \acp{FOV} capturing an \acs{AMR}. The second figure displays a custom function mapping the F1 score to video bitrate.}
    \label{fig:application_picture}
    \vspace{-3.5mm}
\end{figure}

Our scheduler design can be used to serve a variety of use cases and applications that require a minimum application-layer throughput and, in this paper, we demonstrate its effectiveness in the relevant use case of asset tracking in \ac{IIoT}. Specifically, we consider the case where a set of 5G-connected cameras perform asset tracking via computer vision-aided object detection as illustrated in Fig.~\ref{fig:application_picture} (left). Asset tracking requires high \ac{QoS} guarantees as the accuracy of the detection tasks heavily depends on the resolution of video frames received over the 5G uplink channel. As shown in Fig.~\ref{fig:application_picture} (right), the detection accuracy (e.g., the F1-score) depends on the bitrate of the video sent to the computer vision algorithm. The higher the bitrate, the higher the accuracy. Therefore, to achieve a target detection accuracy (or F1-score), the application dictates a minimum application-layer throughput level. How to guarantee this minimum throughput level is not trivial as many cameras might live stream and compete for uplink radio resources, which can result in violations of the minimum throughput requirement due to lack of radio resources. 
However, spatiotemporal correlations of \ac{IIoT} can be used to prioritize cameras that are either in line of sight with the asset, or are is close proximity. In this way, we can assign a different minimum throughput constraint to each camera based on their location, and serve them accordingly to ensure that the asset can be tracked properly.

To address the requirements and consider the properties of the above scenario, we first propose a variant of the \ac{PF} scheduler that includes parameters that control the priority of each \ac{UE}.
We then present an xApp that tunes these parameters to ensure that each \ac{UE}  achieves a throughput above a desired threshold that meets the detection accuracy requirement. The decision made by the xApp is based on knowledge of how the empirical joint \ac{CCDF} of \acp{UE} throughput vary as a function of the scheduler's parameters. We then map application layer requirement of the object detection use case to throughput requirements, to optimize the \ac{RAN} accordingly. 

We evaluate our xApp in an end-to-end O-RAN testbed using \ac{OAI} \cite{KALTENBERGER2020107284} and the \ac{near-RT} \ac{RIC} from OpenRAN Gym \cite{9814869}, which is based on the \ac{OSC} \ac{RIC}.
We present results that demonstrate the effectiveness of our approach in satisfying the minimum throughput requirement in $94.9 \%$ of cases, and show a 33\% improvement with respect to the baseline \ac{PF} scheduler adopted in \ac{OAI}. 
We show that by prioritizing cameras (i.e., users) streaming video to a computer vision-aided asset tracking server, we improve the F1 score by up to 37.04\%, resulting in higher object detection accuracy. 


\section{System Model and Design}\label{sec:system_model}
In this section, we first introduce the system model and describe the use case addressed in this work. Then, we present our scheduler design and show how it can be controlled by an xApp to deliver minimum performance requirements. 

\subsection{Use Case \& System Model}\label{subsec:UC_SM}
Our primary objective is to prioritize certain \acp{UE} to ensure they are guaranteed a minimum \ac{UE}-specific application-layer throughput level. Although the algorithm we illustrate in this paper is general and can be applied to several use cases, for the sake of illustration we focus on the case of asset tracking in industrial surveillance video via object detection, e.g., in the case of \acp{AMR} monitoring. The scenario is illustrated in Fig.~\ref{fig:application_picture} and involves a set of cameras transmitting video over 5G to an application server hosting computer vision-aided object detection services. It has been shown that guaranteeing a minimum bitrate is essential to achieve high accuracy (i.e., F1 scores) in object detection tasks because detection accuracy heavily relies on video quality. Specifically, the lower the compression, the higher the quality and accuracy, but also the higher the required video bitrate~\cite{9959476}. 

The application server processes video streams from each camera to identify \acp{AMR}, and determine which cameras are in line of sight with the \acp{AMR}. By analyzing this video data, the system can identify potential safety risks, as the cameras provide valuable information about the surrounding environment and any hazardous situations involving \acp{AMR}. 
It is worth mentioning that when a camera detects the target, it is likely that the target will either remain in the same area, or move to the \acf{FOV} of adjacent cameras, but will not appear in the \ac{FOV} of cameras farther away. 
We can leverage this spatial information to improve safety by guaranteeing high detection accuracy by computing resource allocation policies that maximize the F1 score and prioritizing streams from cameras likely to be directly pointing at the \ac{AMR}.

\begin{figure}[t!]
    \centering
    \includegraphics[width=0.8\columnwidth]{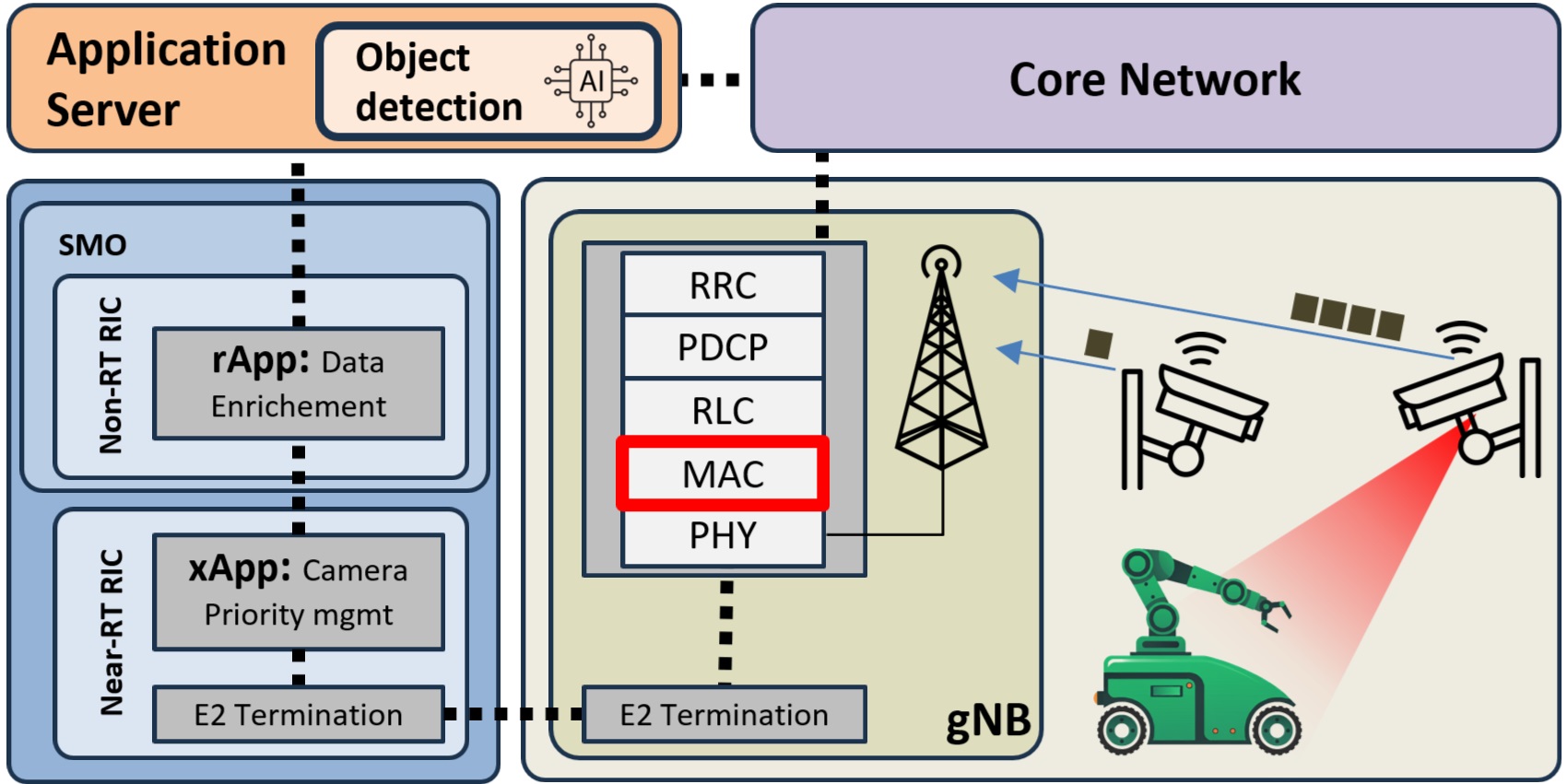} 
    \caption{Proposed setup: 5G-aided object detection use case. Uplink video transmissions from cameras in line of sight with the target are being prioritized via an xApp that adjusts scheduling parameters.}
    \vspace{-4mm}
    \label{fig:architecture_generic}
\end{figure}

The O-RAN architecture is the ideal candidate to enable this prioritization by combining application layer data and information with the ability to reconfigure the RAN to optimize performance. The solution we propose and our scenario are shown in Fig.~\ref{fig:architecture_generic}, where a gNB provides connectivity to a set of \acp{UE}, which, in the selected use case, are cameras. These communicate with an application server directly connected to the \ac{UPF} of the \ac{CN}. The \ac{DU}, which contains the custom scheduler, is connected to the \ac{near-RT} \ac{RIC} via the E2 interface. The scheduler exposes metrics and parameters to the \ac{near-RT} \ac{RIC} through a custom E2 \ac{SM} specifically designed for this purpose. 
This enables access to the implementation-specific scheduler parameters, which would be difficult to handle with a standard \ac{SM}. On the \ac{near-RT} \ac{RIC}, a single xApp is deployed to tune the scheduler, with the goal of optimizing throughput by prioritizing cameras in close proximity with the target. The application server communicates with rApp, providing information regarding the asset tracking task. The rApp, responsible for data enrichment, maps this information into throughput requirements, which are then supplied to the xApp.
\vspace{-1.5mm}
\subsection{Proposed Scheduler}\label{subsec:custom_scheduler}
As of today, the majority of schedulers used commercially are based on optimized variations of the standard \acf{PF} scheduler~\cite{851593, 7515184, 9771597}, which seeks to balance between network throughput and fairness.
This is achieved via the computation of coefficients for each \ac{UE}, based on static weights which are identical for all the \acp{UE} and define the tradeoff between throughput and fairness~\cite{1543653}.

Thanks to its established presence in the market and in the literature, and to showcase a scenario representative of real-world use cases, in this paper we consider \ac{PF} as the reference scheduler. More specifically, since in our use case we need to prioritize each camera differently based on their ability to capture the target, our goal is to control \ac{PF} behavior by computing a different set of weights for each camera (i.e., \ac{UE}). To achieve this, our variation of the \ac{PF} scheduler presents multiple tunable parameters, one per \ac{UE}, which allow us to dynamically change the behavior of the scheduler. Specifically, we introduce the priority coefficient $\gamma_i$ defined as follows:
\begin{equation}
    \gamma_i[n]  = \frac{\theta_i[n]}{D_i[n]}, \quad i \in \{1, \dots, N\},\label{eq:pf_new}
\end{equation}
where $N$ represents the number of \acp{UE}, $i$ refers to each individual \ac{UE}, and $n$ is the current time step. The priority coefficient for the $i$-th \ac{UE} is computed as the ratio between the requested throughput $\theta_i$ of that \ac{UE} and its coefficient $D_i[n]$. Regarding the numerator, the higher the throughput required by an \ac{UE}, the higher its priority: this lead to increasing network throughput. The term $D_i[n]$ is computed as:
\begin{equation}
    D_i[n] = ((1 - \alpha) \cdot D_i[n - 1] + \alpha \cdot \phi_i[n])^{\beta_i},\label{eq:updated_denominator}
\end{equation}
where $\beta_i \in [0, 1]$ is a tunable parameter which allows to adjust how the scheduler allocates resources, and $\phi_i[n]$ is the current throughput of the $i$-th \ac{UE}. $D_i[n]$ represents the amount of data rate provided to the $i$-th \ac{UE} over time. To study the meaning of this term, let us assume $\beta_i = 1~\forall i$: the coefficient is now computed as an exponential moving average, which smooths the time series of data rate values. The raw data sequence is denoted as $\{\phi_i\}$, and the smoothed sequence is represented as $\{D\}$. How much smoothing is applied to the sequence is determined by the smoothing coefficient $\alpha  \in [0, 1]$. The denominator, which measures the amount of throughput allocated to an \ac{UE}, is  related to fairness: the higher the amount of resources provided to an \ac{UE} in the past, the lower its priority in the current scheduling decision. The effect of varying $\beta_i$ is as follows: with all other parameters held constant, a smaller value of $\beta_i$ results in higher priority coefficient for the $i$-th \ac{UE}, effectively allowing to increase the priority, and specifically provide more resources, to that \ac{UE}. Notably, when $\beta_i = 1~\forall i$, the scheduler is a standard \ac{PF} scheduler.

Thanks to these parameters that enable to balance fairness with prioritization, the scheduler offers the necessary control knobs for an xApp to fine-tune the scheduler. How to achieve this is described below.

\subsection{Controlling Scheduler Weights via an xApp}\label{sec:xapp_design}

The xApp employs a data-driven approach to control the scheduler. We collect data to understand the impact of the scheduler’s parameters on throughput. This allows us to identify the beta values that statistically guarantee, with a specified probability, that the throughput for each \ac{UE} exceeds its required threshold. As a result, the xApp is designed to ensure that the throughput of each \ac{UE} remains above a specified target—unique for each \ac{UE}—with a certain probability.

The design of our xApp requires four steps, which are summarized here and detailed below: 
\begin{enumerate} 
    \item Gather data from experiments by varying beta parameters $\bm{\beta} = (\beta_1, \dots, \beta_N)$. 
    \item For each set of beta parameters, fix $\bm{\beta}$ and compute the empirical probability that the throughput for each \ac{UE} exceeds a specified threshold. 
    \item Identify the $N$-dimensional level set corresponding to the desired \ac{QoS} requirement. 
    \item Aggregate all the level sets and create a hash table that maps throughput requirements to corresponding $\bm{\beta}$ parameters. 
\end{enumerate}

In Step (1), we vary $\bm{\beta}$ and generate traffic from the \acp{UE}. We consider a finite set of parameters $\bm{\beta}$ defined as follows:
\begin{equation}
    \begin{aligned}
        \bm{\beta} = &(\beta_1, \dots, \beta_N) \in \\
        &\{ (x_1, \dots, x_N) \mid x_i \in B \text{ for } i = 1, \dots, N\}
    \end{aligned}\label{eq:set_of_betas}
\end{equation}
\noindent
where $B = \{b_1, \dots, b_m\}$ and $b_j  \in [0, 1]$.

In Step (2), for each set of $\bm{\beta}$ values, we compute the empirical probability that each \ac{UE}'s throughput exceeds a target value, thereby estimating the joint \ac{CCDF}, defined as: \begin{equation} 
    \bar{F}_{\bm{\Phi}}(\bm{\phi}) = P(\Phi_1 \geq \phi_1, \dots, \Phi_N \geq \phi_N) 
\end{equation} To simplify notation, we denote the estimate of the joint \ac{CCDF} as $S(\bm{\phi})$, where $\bm{\phi}$ refers to the throughput vector of the \acp{UE}. Throughput is calculated over a time window $T_A$.

In Step (3), once we have $S(\bm{\phi})$, we find the $N$-dimensional level set corresponding to a specific value $q$ as follows: 
\begin{equation} 
I_q = \{\bm{\phi} \mid S(\bm{\phi}) = q\} \end{equation} 
This gives us the set of throughput points such that the probability of exceeding the throughput set by each point is $q$.

In Step (4), we aggregate all the data from the level sets into a hash table which maps the throughput points (keys) from the level sets to their corresponding $\bm{\beta}$ values (values). This hash table is used by the xApp to map any minimum throughput requirement to the $\bm{\beta}$ values that satisfy the requirement.

When the xApp receives throughput requirements, it first filters the keys of the hash table, removing all entries where the throughput points (keys) are lower than the required throughput. After this initial filtering, if multiple entries remain, we apply two additional filters based on expert knowledge: \begin{itemize} 
    \item If the throughput requirements for certain \acp{UE} are equal, the corresponding $\beta_i$ values must also be equal. 
    \item Since lower $\beta_i$ value corresponds to higher priority (and thus higher throughput) for $i$-th \ac{UE}, we sort the throughput requirements in ascending order and ensure that the corresponding $\beta$ values are in descending order. 
\end{itemize} 
These filters reduce the set of potential $\bm{\beta}$ parameters to a small subset, from which one is selected at random.

\subsection{Applying the Scheduler to the Use Case}\label{subsec:use_case_xapp_optimization}

Since higher video bitrates correspond to higher F1 scores, we can increase the throughput of users to improve their F1 scores using the xApp defined above. Specifically, the data enrichment rApp, which collects information from the application server, can map application requirements to throughput requirements, which are subsequently provided to the xApp. To guarantee a high F1 score for cameras framing the target, we need to guarantee a minimum throughput level to such cameras. The work in \cite{9959476} presents models that map F1 scores to required video bitrates. 

The relationship between F1 score and datarates can be modeled as power functions in the form 
\begin{equation}
    y~=~a~\cdot~x^b~+~c,
\end{equation}
\noindent
whose parameters can be configured to represent diverse relationships bewteen F1 score and datarates based on specific image recognition tasks. In the following, we show how different configurations affect prioritization and network performance.

\section{Prototype Implementation}

We leverage the Colosseum testbed to evaluate the performance of the setup presented in Sec.\ref{sec:system_model}. Both gNB and \acp{UE} are implemented using \ac{OAI}, while the channels are emulated through Colosseum \ac{MCHEM}. In \ac{OAI}, the default \ac{PF} scheduler is implemented considering the \ac{TBS} instead of throughput, as noted in \cite{9771597}. Therefore, the $\theta_i$ and $\phi_i$ referred to in Sec.~\ref{subsec:custom_scheduler} are related to \ac{TBS} rather than throughput.
The \ac{near-RT} \ac{RIC} used is based on the one presented in \cite{9814869}, which adapts the \ac{OSC} \ac{near-RT} \ac{RIC} to run on Colosseum. Further, we modified the E2 interface to connect the \ac{near-RT} \ac{RIC} with the \ac{OAI}-based gNB.

In our experiments, we simulate camera feed traffic using iPerf3. Then, to assess the benefits of xApp-based scheduling control in our object detection use case, the xApp adjusts dynamically scheduler weights for each \ac{UE} so as to prioritize those cameras that are in line of sight with the target, ensuring that the bitrate of the video stream is high enough to detect the target with a minimum accuracy level. This information is provided by an rApp that maps the throughput required by a simulated application server using computer vision for target detection to the desired detection accuracy level.
We modeled the scenario with four cameras, each positioned at the corners of a sector of an indoor factory environment. When a camera is actively framing a target, it is assigned the highest priority, while the camera opposite to it receives the lowest priority. The two remaining cameras are assigned an intermediate priority. We test the system according to the method defined in Sec.~\ref{subsec:use_case_xapp_optimization}.


\section{Experimental Evaluation}\label{sec:system_performance1}
In this section, we present results obtained on our prototype. First, we provide insights on the behavior of the proposed scheduler, then, we analyze the performance of the system when an xApp interacts with the scheduler to guarantee a minimum throughput to each \ac{UE}. Finally, we evaluate the system performance in the context of object detection in surveillance videos for industrial application scenarios.

\begin{figure}[t!]
    \centering
    \includegraphics[width=\columnwidth]{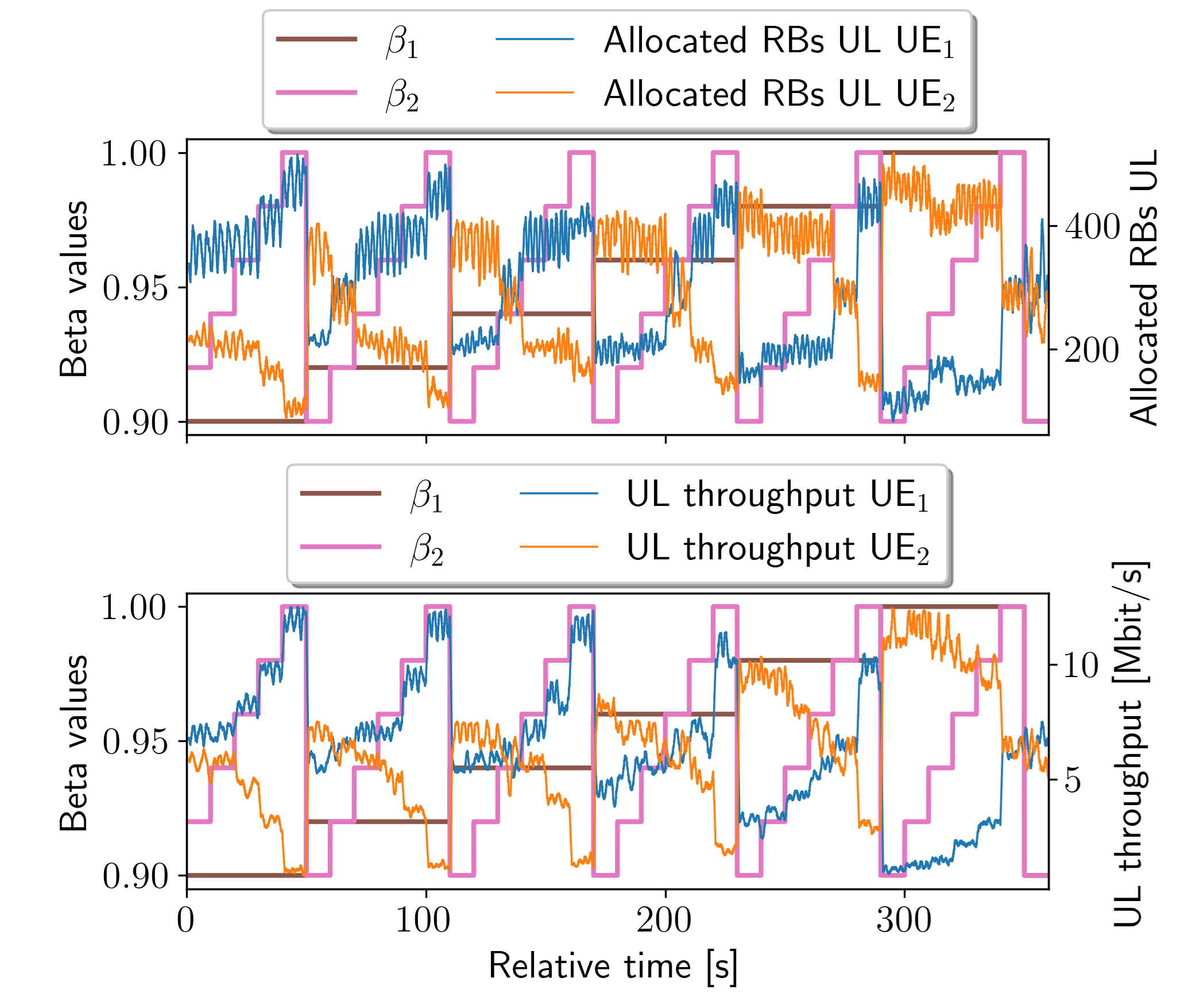} 
    \vspace{-4mm}
    \caption{Performance of the custom scheduler illustrating how the number of allocated \acsp{RB} and the throughput change as a function of the beta parameters.}
    \label{fig:scheduler_2ue}
     \vspace{-3mm}
\end{figure}

\subsection{Impact of Scheduler on UE performance}
To understand how the proposed scheduler affects \ac{UE} performance, we perform an experiment with uplink traffic from 2 \acp{UE} and manually vary $\beta_1$ and $\beta_2$ of eq.~\eqref{eq:updated_denominator}. Similarly to what we did in eq.~\eqref{eq:set_of_betas}, each $\beta_i$ can assume values in $B = \{0.9, 0.92, 0.94, 0.96, 0.98, 1\}$, and we test all their possible combinations. Results are presented in Fig.~\ref{fig:scheduler_2ue}. In the top plot, we observe how the number of allocated \acp{RB} changes with $\beta_i$. As $\beta_i$ decreases for a \ac{UE}, the number of \acp{RB} allocated to that \ac{UE} increases, while the number of \ac{RB} allocated to the other \ac{UE} decreases. This behavior aligns with eq.~\eqref{eq:pf_new} which suggests that each $\beta_i$ influences all the \acp{UE} in the system. Since the scheduler manages all the \acp{UE} collectively, increasing the priority of one \ac{UE} inevitably lowers priority of other \acp{UE}. The bottom graph, shows a similar behavior for the uplink throughput of the \acp{UE}.

\subsection{Minimum Throughput Guarantee via xApp}
We now analyze the performance of the system when the scheduler weights are optimized by the xApp aiming to guarantee a minimum throughput for each \ac{UE}. In this experiment, we set $N = 4$, $B = \{0.8, 0.85, 0.9, 0.95, 1\}$, $q = 0.99$, and $T_A = 50$ ms.
In Table \ref{tab:performance_table}, we display the throughput values provided as input to the xApp (first column) alongside the optimal beta values it determines (second column). We also include the success rate $P_S$ (third column), which represents the ratio of occurrences in which each \ac{UE} meets or exceeds its throughput requirements,
and the improvement $\Delta  P_S$ in the success percentage relative to the default \ac{OAI} \ac{PF} scheduler (fourth column), which we use as baseline. The throughput requirements are determined by testing the throughput of each \ac{UE} in three cases, each representing different conditions:
\begin{itemize}
    \item In Case 1, requirements [3,~0.2,~0.2,~0.2]~Mbit/s, we prioritize one \ac{UE}.
    \item In Case 2, requirements [1.4,~1.4,~0.4,~0.4]~Mbit/s, we give moderate priority to two \acp{UE}, treating them equally.
    \item In Case 3, requirements [2.4,~1.2,~0.4,~0.4]~Mbit/s, we significantly prioritize one \ac{UE}, while giving lower priority to another \ac{UE}.
\end{itemize}
Our system demonstrates improved performance, with an average success percentage of 94.9\% and a 33.0\% improvement with respect to the baseline. The improvement is greater when fairness we require is lower. In particular, Case 1 shows the best improvement due to the significant prioritization of one \ac{UE}. We also observe that, considering the same case, both the success percentage and the improvement vary with the prioritized \acp{UE}. 

\begin{table}[ht]
    \vspace{4mm}
    \centering
    \fontsize{7}{6}\selectfont

    \begin{tabular}{@{}p{0.15cm}|p{0.4cm}p{0.4cm}p{0.4cm}p{0.4cm}|p{0.4cm}p{0.4cm}p{0.4cm}p{0.4cm}|p{0.5cm}|p{0.7cm}@{}}
        \toprule
        & \multicolumn{4}{c|}{\fontsize{8}{7}\selectfont\textbf{Throughput}} & \multicolumn{4}{c|}{\fontsize{8}{7}\selectfont\textbf{Beta}} & \textbf{} & \\ 
        & \multicolumn{4}{c|}{\fontsize{8}{7}\selectfont\textbf{req. [Mbit/s]}} & \multicolumn{4}{c|}{\fontsize{8}{7}\selectfont\textbf{parameters}} & \fontsize{8}{7}\selectfont\textbf{$P_S$} & \fontsize{8}{7}\selectfont\textbf{$\Delta  P_S$}\\ 
        \cline{2-9}
        \rule{0pt}{10pt}& UE1 & UE2 & UE3 & UE4 & $\beta_1$ & $\beta_2$ & $\beta_3$ & $\beta_4$ & & \\ 
        \midrule
        \multirow{4}{*}{\rotatebox{90}{\textbf{Case 1}}} 
        & 3.0 & 0.2 & 0.2 & 0.2 & 0.8 & 0.95 & 0.95 & 0.95 & 98.8 & 38.2 \\
        & 0.2 & 3.0 & 0.2 & 0.2 & 0.95 & 0.8 & 0.95 & 0.95 & 93.2 & 76.4 \\
        & 0.2 & 0.2 & 3.0 & 0.2 & 0.95 & 0.95 & 0.8 & 0.95 & 95.2 & 84.5 \\
        & 0.2 & 0.2 & 0.2 & 3.0 & 0.95 & 0.95 & 0.95 & 0.85 & 99.3 & 75.1 \\ \midrule

        \multirow{6}{*}{\rotatebox{90}{\textbf{Case 2}}}
        & 0.4 & 0.4 & 1.4 & 1.4 & 0.95 & 0.95 & 0.9 & 0.9 & 97.0 & 8.3 \\
        & 0.4 & 1.4 & 1.4 & 0.4 & 0.9 & 0.8 & 0.8 & 0.9 & 97.8 & 17.0 \\
        & 1.4 & 0.4 & 0.4 & 1.4 & 0.85 & 0.95 & 0.95 & 0.85 & 96.5 & 0.8 \\
        & 1.4 & 0.4 & 1.4 & 0.4 & 0.8 & 0.9 & 0.8 & 0.9 & 98.3 & 7.8 \\
        & 0.4 & 1.4 & 0.4 & 1.4 & 0.95 & 0.85 & 0.95 & 0.85 & 98.1 & 12.6 \\
        & 1.4 & 1.4 & 0.4 & 0.4 & 0.8 & 0.8 & 0.9 & 0.9 & 95.6 & 8.2 \\ \midrule

        \multirow{12}{*}{\rotatebox{90}{\textbf{Case 3}}}
        & 2.4 & 1.2 & 0.4 & 0.4 & 0.85 & 0.9 & 0.95 & 0.95 & 93.6 & 14.9 \\
        & 1.2 & 2.4 & 0.4 & 0.4 & 0.85 & 0.8 & 0.95 & 0.95 & 95.4 & 39.0 \\
        & 1.2 & 0.4 & 0.4 & 2.4 & 0.85 & 0.95 & 0.95 & 0.85 & 95.6 & 29.3 \\
        & 0.4 & 2.4 & 0.4 & 1.2 & 0.95 & 0.85 & 0.95 & 0.85 & 92.7 & 36.9 \\
        & 2.4 & 0.4 & 0.4 & 1.2 & 0.8 & 0.95 & 0.95 & 0.85 & 88.1 & 1.7 \\
        & 1.2 & 0.4 & 2.4 & 0.4 & 0.85 & 0.95 & 0.8 & 0.95 & 93.6 & 53.4 \\
        & 2.4 & 0.4 & 1.2 & 0.4 & 0.8 & 0.95 & 0.85 & 0.95 & 92.5 & 8.7 \\
        & 0.4 & 2.4 & 1.2 & 0.4 & 0.9 & 0.8 & 0.8 & 0.9 & 86.8 & 33.0 \\
        & 0.4 & 1.2 & 0.4 & 2.4 & 0.95 & 0.85 & 0.95 & 0.85 & 98.1 & 38.1 \\
        & 0.4 & 0.4 & 2.4 & 1.2 & 0.95 & 0.95 & 0.8 & 0.9 & 96.3 & 56.5 \\
        & 0.4 & 1.2 & 2.4 & 0.4 & 0.95 & 0.85 & 0.8 & 0.95 & 90.3 & 53.3 \\
        & 0.4 & 0.4 & 1.2 & 2.4 & 0.95 & 0.95 & 0.85 & 0.8 & 95.3 & 31.6 \\ \midrule

        \multicolumn{9}{c|}{\textbf{AVERAGE}} & 94.9 & 33.0 \\
        \bottomrule
    \end{tabular}
    \caption{Throughput requirements for the xApp and the corresponding beta parameters, along with the success percentage $P_S$ and improvements $\Delta P_S$ compared to the baseline.}
    \label{tab:performance_table}
    \vspace{-2mm}
\end{table}

\begin{figure}[t!]
  \centering \includegraphics[width=0.88\columnwidth]{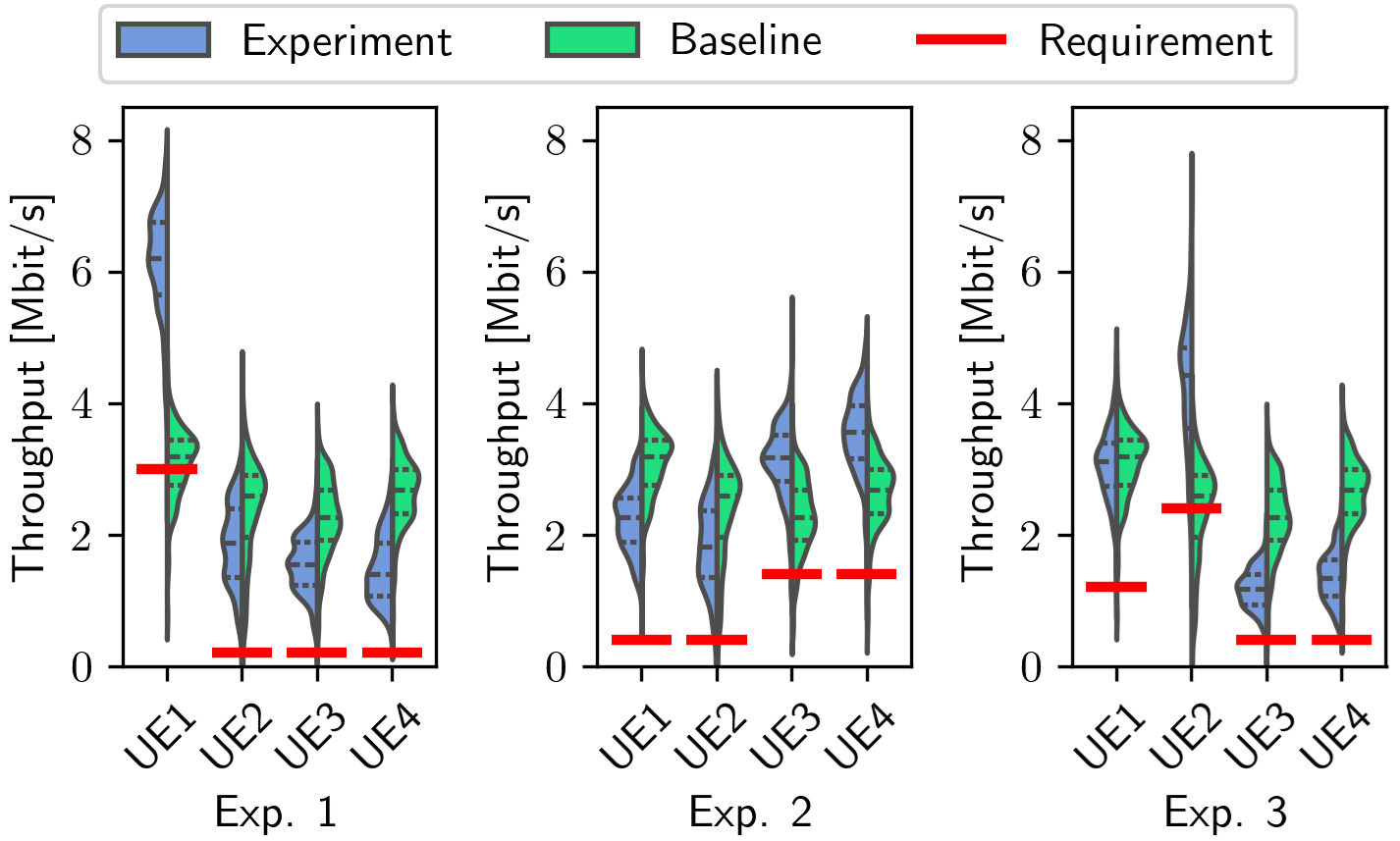} 
    \caption{Violin plot showing uplink throughput for each \ac{UE}, with the xApp's requirements represented for each \ac{UE} with red lines. 
    Baseline performance is provided by the default \ac{OAI} scheduler.}
    \label{fig:violin_plot}
    \vspace{-.4cm}
\end{figure}

In Fig.~\ref{fig:violin_plot}, we illustrate how varying requirements affect \acp{UE} by showing the throughput distribution for each \ac{UE}.
We test one sequence for each case described above, using the default \ac{OAI} scheduler as baseline. The plots illustrate how the throughput distribution for each \ac{UE} varies consistently with the specified requirements. 
It is important to note that the xApp adjusts the beta parameters to guarantee throughput values that satisfy the requirements of all \ac{UE} simultaneously; and does not focus on maximizing throughput for individual \acp{UE}.
Furthermore, we observe that, despite prioritizing throughput for each \ac{UE}, the performance slightly varies among \acp{UE} with equal parameters. This variation is due to the non-ideal behavior of \ac{OAI}, which is also observed with the default \ac{PF} scheduler.

\begin{figure}[t!]
\setlength\abovecaptionskip{-5pt}
    \centering
    \vspace{-1mm}
    \begin{subfigure}[t]{0.48\columnwidth}
        \setlength\abovecaptionskip{-0pt}
        \centering
        \includegraphics[width=\textwidth]{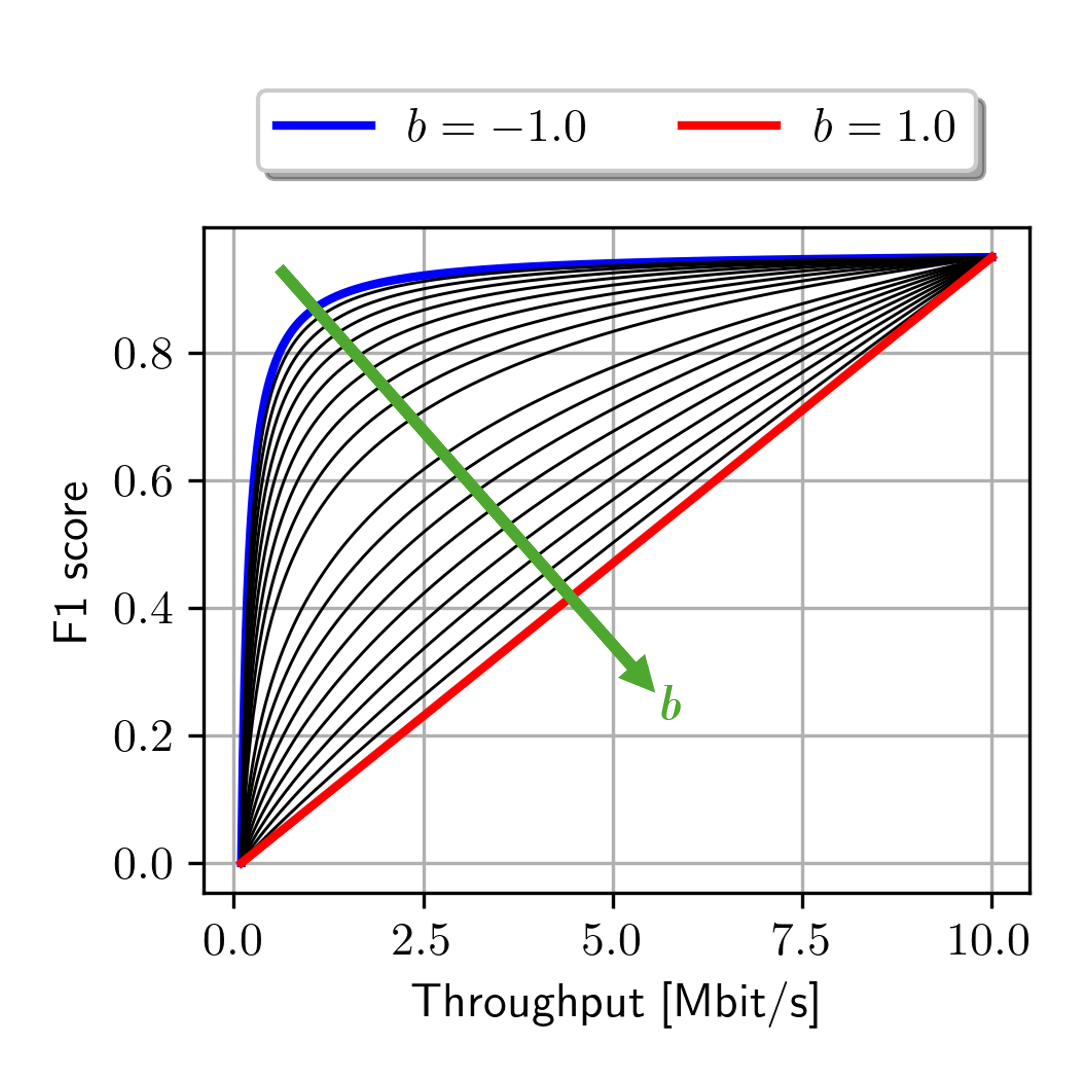}
        \caption{Impact of $b$ in the power function $y = a \cdot x^b + c$, where $x = 0.1$ and $x = 10$ are fixed. $b$ controls the steepness of the curves. Curves for $b = -1$ (blue) and $b = 1$ (red) are highlighted, while intermediate curves represent values of $b \in [-1, 1]$ with $0.2$ increments. The green arrow indicates the direction of increasing $b$.}
        \label{fig:f1_vs_throughput}
    \end{subfigure}
    \hspace{1mm}
    \begin{subfigure}[t]{0.48\columnwidth}
        \setlength\abovecaptionskip{-0pt}
        \centering
        \includegraphics[width=\textwidth]{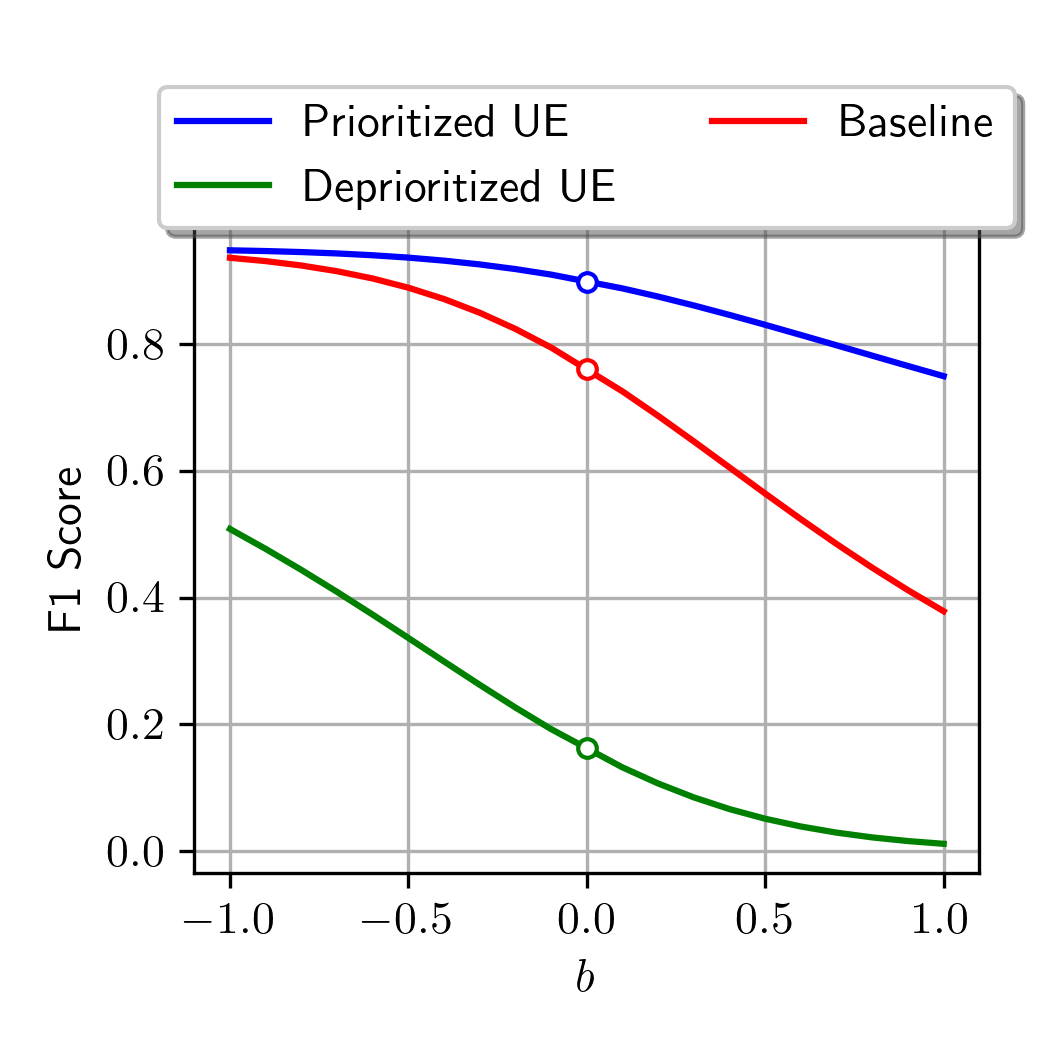}
        \caption{Relationship between the F1 score and the parameter $b$ which controls the shape of the curves, as defined in Fig.~\ref{fig:f1_vs_throughput}. The plot presents three cases: prioritized user equipment \ac{UE}, deprioritized \ac{UE}, and a baseline scenario using the default \ac{OAI} \ac{PF} scheduler. The curves are not defined in $b=0$. \\ \\ }
        \label{fig:prova2}
    \end{subfigure}
    \caption{Impact of the $b$ parameter on the mapping of F1 score to throughput, and its effect on F1 score based on the priority assigned to \acp{UE}.}
    \label{fig:application_performance}
    \vspace{-4.5mm}
\end{figure}

\subsection{Object Detection Optimization via an xApp}
In the next experiment, the xApp configures the scheduler to optimize object detection accuracy of moving targets in surveillance video within an industrial scenario. Considering the spatial correlation of this use case,
we aim to provide 2 Mbit/s to the camera framing the target, 1 Mbit/s to the adjacent cameras, and 0 Mbit/s to the camera that is further away. 

As discussed in Sec.~\ref{subsec:use_case_xapp_optimization}, we model the relationship between the F1 score and throughput using a power function. We fix the throughput that results in no detection at $0.1$ \text{Mbit/s} and set the F1 score to $0.95$ when the throughput is $10$ \text{Mbit/s}. With these constraints, varying the parameter $b$ alters the shape of the curve. The results are shown in Fig.~\ref{fig:application_performance}: the left plot illustrates how the curve changes with $b$, while the right plot shows the F1 score variation with $b$ for three types of users---the prioritized user, the deprioritized user, and the baseline (default \ac{OAI} scheduler). Notably, the improvement of the prioritized user over the baseline increases as $b$ increases, which is explained by the changes in slope with $b$ in the left plot.

In Table \ref{tab:performance_table_KPI}, we report the system performance.
The average F1 score improves by between $1.17\%$ and $37.04\%$, depending on the $b$ parameter
for the camera currently detecting the target. Conversely, we observe a  
degradation ranging from $-36.6\%$ to $-60.1\%$ for the farthest camera. The average processing time of the xApp---the time from when we set the requirement to when we obtain the beta parameters as output---is $1.22$ ms. Additionally, the control loop time, defined as the duration from when we generate the throughput requirements to when the gNB generates the first log containing the updated beta parameters, is $36.67$ ms. We also test the control latency, defined as the time between generating the throughput requirements and collecting the first throughput sample that meets those requirements, which averages $85.33$ ms. The minimum value is $T_A = 50$ ms, since we must wait $T_A$ before computing the first throughput sample. Given that the average control latency is nearly equal to the control loop time plus $T_A$, we can conclude that the parameters set by our xApp are immediately effective on the network, indicating the absence of dynamic effects.

\begin{table}[h!]
\scriptsize
\centering
\begin{adjustbox}{max width=\textwidth}
\begin{tabular}{|l|l|}
\hline
\textbf{KPI} & \textbf{Performance} \\
\hline
xApp processing time & $1.22 \pm 0.05$ ms\\
\hline
Control loop time & $36.67 \pm 6.40$ ms\\
\hline
Control latency & $85.33 \pm 13.15$ ms\\
\hline
F1 score improvement prioritized camera & $[1.17\%,~37.04\%]$ \\
\hline
F1 score deterioration deprioritized camera & $[-36.6\%,~-60.1\%]$ \\
\hline
\end{tabular}
\end{adjustbox}
\caption{Performance analysis of xApp-based scheduling control in the object detection use case.}
\label{tab:performance_table_KPI}
\vspace{-3mm}
\end{table}

\section{Conclusions}


In this paper, we proposed an O-RAN-based solution for fine-tuning 5G \ac{RAN} schedulers to meet specific performance requirements. We addressed the complex challenges of delivering the necessary performance for verticals served by 5G and future 6G systems, among which \ac{IIoT} is one of the most demanding.
%

Our solution consists of a tunable and enhanced version of \ac{PF} scheduler.
Parameters are exposed to an xApp that tunes them to satisfy network objectives.
%
Specifically, we designed an xApp that tunes the scheduler to achieve a joint minimum throughput for each \ac{UE} by utilizing information on their throughput distribution.
%
We implemented the proposed scheduler in \ac{OAI}, utilizing the \ac{OSC} near-RT \ac{RIC} to host our custom xApp, and demonstrated the effectiveness of our approach on the Colosseum testbed, achieving an average success percentage of 94.9\% in meeting throughput targets, which represents a 33\% improvement over the default \ac{OAI} scheduler. 
Focusing on the IIoT use case of object detection in surveillance videos within industrial scenarios, by leveraging information about the application we are serving, we optimized the \ac{RAN} to enhance the application performance, increasing the F1 score of the camera framing the target by up to 37.04\%. 


\balance
\bibliographystyle{IEEEtran}
\bibliography{IEEEabrv,StringDefinitions,references}

\end{document}